\title{On geodesic dynamics in deformed black-hole fields}

\author{O. Semer\'ak\footnote{Email: oldrich.semerak@mff.cuni.cz, URL: utf.mff.cuni.cz/en/index.html}\\
        Institute of Theoretical Physics,\\
        Faculty of Mathematics and Physics,\\
        Charles University in Prague,\\
        Czech Republic\\[5mm]
        and
        P. Sukov\'a\footnote{Email: psukova@cft.edu.pl, URL: http://www.cft.edu.pl/en/index.php}\\
        Center for Theoretical Physics,\\
        Polish Academy of Sciences, Warsaw,\\
        Poland}

\date{\today}

\documentclass[11pt,english]{article}
\usepackage[]{graphicx}
\usepackage{caption}
\usepackage[linktocpage=true]{hyperref}

\begin{document}
\maketitle

\begin{abstract}
``Almost all" seems to be known about isolated stationary black holes in asymptotically flat space-times and about the behaviour of {\em test} matter and fields in their backgrounds. The black holes likely present in galactic nuclei and in some X-ray binaries are commonly being represented by the Kerr metric, but actually they are not isolated (they are detected only thanks to a strong interaction with the surroundings), they are not stationary (black-hole sources are rather strongly variable) and they also probably do not live in an asymptotically flat universe. Such ``perturbations" may query the classical black-hole theorems (how robust are the latter against them?) and certainly affect particles and fields around, which can have observational consequences. In the present contribution we examine how the geodesic structure of the static and axially symmetric black-hole space-time responds to the presence of an additional matter in the form of a thin disc or ring. We use several different methods to show that geodesic motion may become chaotic, to reveal the strength and type of this irregularity and its dependence on parameters. The relevance of such an analysis for galactic nuclei is briefly commented on.
\end{abstract}

\section{Introduction}

Geodesic structure is a very comprehensive and demonstrative attribute of space(-time). As traversing regions of all possible sizes, geodesics can unveil a local behaviour of a given system as well as tiny tendencies only discernible over an extensive span of time. A default example of the latter are weak irregularities attending a lack of a complete set of integrals of the motion. Actually, the long-term geodesic dynamics is a suitable tool how to detect, illustrate, evaluate, classify and compare different deviations of a chosen system from a certain simple, ``regular" ideal.
In mathematics and physic, such an ideal is represented by linear systems (the finite-dimensional in particular). However, within the last 150 years it has become clear that even in these highly abstract fields the linear systems represent just marginal tips within a vast non-linear tangle which is typically prone to ``irregularities" and which can display ``chaotic" behaviour even in rather simple settings.

The modern {\it theory of chaos} was apparently inspired by Henri Poincar\'e's treatment of a three-body system, and our interest will also focus on systems driven by {\em gravitational} interaction in this contribution. Specifically, we consider a simplest possible gravitational centre, the Schwarzschild black hole, and study how the time-like geodesics in its field respond on perturbation due to the presence of a very simple additional source, namely a static and axially symmetric thin annular disc or ring. Such a topic has an extra {\it attraction} in general relativity, due to the latter's non-linearity. Besides that, as reminded in \cite{SemerakS-10}, it is doubly apropos to seek for chaos around black holes, because in ancient Greek {\sc chaos} meant a gaping bottomless void, where everything falls endlessly\dots A more sophisticated reason was articulated nicely by \cite{CornishF-97}:
``\dots even the most pristine black-hole space-time harbours the seeds of chaos in the form of isolated unstable orbits. A small perturbation causes these unstable orbits to break out and infect large regions of phase space. Note that the experience with Newtonian systems is very misleading. For example, the Kepler problem has more integrals of motion than are needed for integrability. Keplerian systems are thus impervious to small perturbations. In contrast, black hole space-times are at the edge of chaos, just waiting for the proverbial butterfly to flap its wings."

However, the space-time of a black hole enclosed by a disc or ring has not been chosen only for theoretical reasons: the black holes probably present in galactic nuclei and in some X-ray binaries are supposed to be interacting with their surroundings through disc accretion, and, besides that, galactic nuclei are also typically encircled, at larger distance, by tori of colder molecular gas and dust (called circum-nuclear rings). These structures (at least the inner accretion disc) are assumed to be rather lightweight with respect to the central black hole, and theoretical models actually treat them as test (not contributing to the gravitational field at all). Such an approximation is certainly adequate concerning the potential (metric) and its gradient (field), but may not hold for higher derivatives (curvature). However, the curvature terms are important for stability of the motion, so one can expect that a gravitating matter would assume a {\em different} configuration than a test one. This could be important in the physics of accretion discs, as the latter e.g. depends crucially on position of the disc inner radius (approximated by the innermost stable circular orbit); this in turn plays a key role in estimating the black-hole spin from observations. The issue was pointed out already by \cite{AbramowiczCSW-84} on Newtonian grounds and later confirmed in analysis of the sequences of exact space-times describing the fields of a black hole with concentric thin discs (e.g. \cite{Semerak-03,Semerak-04} and references therein). It turned out, in particular, that the properties of circular motion in the disc plane are indeed altered due to the orbiting-matter own gravity, for example, the frequencies of epicyclic oscillations indicate that the disc may remain stable closer to the horizon, whereas it rather inclines to instability at larger (intermediate) radii (e.g. \cite{ZacekS-02}).

It should be stressed right away that a test body orbiting in a black-hole field has various other reasons why to behave in a chaotic way. Even if leaving aside the most probable perturbation, namely due to a {\em mechanical} interaction with the ambient medium, and focusing only on {\em gravitational} effects, the dynamics depends on how the body is described (whether it is point-like, i.e. without structure, or endowed with more multipoles than just mass), on whether/how it is influenced by incident gravitational and electromagnetic waves, and on whether/how the gravitational and electromagnetic emissions of the body and corresponding reactions on itself are taken into account. Besides that, in reality the body also feels back reaction due to its own gravity (it is not strictly test). Finally, one has to take care of what kind and amount of chaos is added by the numerical code alone: numerical truncations are easily under control for regular orbits along which a sufficient number of quantities (a ``complete set of integrals") remains constant, but for chaotic trajectories such a full check is not at hand. It is really confirmed by experience that even with just slightly modified code it is almost impossible to reproduce {\em accurately} a given long-term chaotic evolution. This urges caution in interpreting the results of chaotic-dynamics modelling and suggests not to trust it uncritically down to quantitative details.

In such an area, it is of particular importance to use several independent methods (and, if possible, also different basic codes for integration of the equations of motion) and compare their outcomes. The rapid progress that the field of chaotic dynamics have been experiencing since the middle of the 20th century has indeed yielded a number of techniques which, starting from the beginning of 1990s, have been also applied to general relativistic systems. One can roughly divide them into two groups: those which require the knowledge of the system's dynamics (evolution equations) and those which can manage with just series of values (time series of experimental data, for example). Of the first group, we employed the Poincar\'e surfaces of section, the Lyapunov exponents and two other similar indicators (abbreviated as FLI and MEGNO) and also followed the evolution of the so called latitudinal action (given by the component of four-momentum which is not bound by any constant of motion). Concerning the second group of methods, we drew the power spectrum of a certain dynamical variable (``vertical" position of the particle in our case) and subjected the motion to two variants of recurrence analysis, one focused on directions in which the orbits recurrently traverse ``pixels" of a prescribed phase-space grid and the other based on recurrences to the chosen cells themselves.

Below we first describe our system in more detail. Then, in section \ref{chaos-methods}, we briefly outline methods we have used to analyse the regime of geodesic dynamics in the chosen black-hole--disc/ring field, and review basic observations they provide. Comments concerning astrophysical relevance of the results and possible further extensions of the work are given in Concluding remarks. We will not go into details, especially not concerning the theory of chaos itself and the diagnostic methods used, focusing mainly on results and their interpretation. For a thorough account, please, see the papers \cite{SemerakS-10,SemerakS-12,SukovaS-13} and references therein. Let us add that we use geometrized units in which $c=1$ and $G=1$ and the metric ($g_{\mu\nu}$) signature ($-$+++).

\section{Static and axially symmetric metrics for a black hole surrounded by discs or rings}

Due to the non-linearity of Einstein equations, the fields of multiple sources are very difficult to find in general.
In some situations, however, the equations simplify to a form which permits to solve at least a certain part of the problem in a Newtonian manner (namely, certain metric components superpose linearly). One very important example of such a setting is a static and axially symmetric case. Choosing the parameters of these symmetries as the time and azimuthal coordinates, $t$ and $\phi$, the {\em vacuum} metric can always be written then in the Weyl form
\begin{equation}
  {\rm d}s^2=-e^{2\nu}{\rm d}t^2+\rho^2 e^{-2\nu}{\rm d}\phi^2
             +e^{2\lambda-2\nu}({\rm d}\rho^2+{\rm d}z^2)
\end{equation}
involving just two unknown functions $\nu$ and $\lambda$ depending only on coordinates $\rho$ and $z$ (cylindrical-type radius and ``vertical" axis) which cover the meridional surfaces. The $\nu$ function represents Newtonian gravitational potential and is given by Laplace equation, hence it superposes linearly. The other function $\lambda$ represents non-Newtonian part of the problem. It is given by line integration of the derivatives of $\nu$, reducing to zero along the vacuum parts of the axis; it does not superpose linearly and except some special cases it has to be found numerically.

In general relativity, the motion has 3 degrees of freedom in general, because four-velocity $u^\mu$ is always constrained by normalization $g_{\mu\nu}u^\mu u^\nu=-1$. In the above space-times, we have two independent constants of {\em geodesic} motion thanks to the stationarity and axial symmetry, namely energy and angular momentum with respect to infinity per unit particle mass, ${\cal E}=-g_{tt}u^t$ and $\ell=g_{\phi\phi}u^\phi$. In contrast to the space-times of {\em isolated} stationary black holes, there is no irreducible Killing tensor and consequently no other independent conserved quantity (the so-called Carter constant quadratic in four-velocity), which implies that the geodesic motion may become chaotic.

We have been interested in time-like geodesic dynamics in the field of a Schwarzschild black hole surrounded, in a concentric way, by an annular thin disc or ring. Specifically, we considered one of the discs (mainly the first one) of the counter-rotating Morgan-Morgan family, inverted (Kelvin-transformed) with respect to their rim (see e.g. \cite{LemosL-94,Semerak-03}),\footnote
{We also checked that the results are similar for discs of the family with power-law density profile \cite{Semerak-04} which are however more demanding computationally.}
and also the Bach-Weyl solution for a circular ring (e.g. \cite{SemerakZZ-99a,DAfonsecaLO-05}).
The Newtonian surface densities of the inverted Morgan-Morgan (iMM) discs are
\begin{equation}  \label{densitydisc}
  w_{\rm iMM}^{(m)}=
  \frac{2^{2m}(m!)^2}{(2m)!\,\pi^2}\,\frac{{\cal M}b}{\rho^3}
  \left(1-\frac{b^2}{\rho^2}\right)^{\!m-1/2}
\end{equation}
and their fields are described by the potentials
\begin{equation}  \label{nu(m)disc}
  \nu^{(m)}_{\rm iMM}=
 -\frac{2^{2m+1}(m!)^2}{\pi}\,\frac{{\cal M}}{b}\;
  \frac{\sum\limits_{n=0}^{m}
        C^{(m)}_{2n}\;
        {\rm i}Q_{2n}\!\left(\!\frac{{\rm i}|y|}{\sqrt{x^2+1-y^2}}\!\right)
        P_{2n}\!\left(\!\frac{x}{\sqrt{x^2+1-y^2}}\!\right)}
       {\sqrt{x^2+1-y^2}} \;,
\end{equation}
where the coefficients read
\[C^{(m)}_{2n}=\frac{(-1)^n (4n+1)(2n)!(m+n)!}
                    {(n!)^2 (m-n)!(2m+2n+1)!}
  \quad\quad (n\leq m) \,,\]
$P_{2n}(y)$ are Legendre polynomials and $Q_{2n}({\rm i}x)$ are Legendre functions of the second kind,
${\cal M}$ and $b$ are the disc mass and Weyl inner radius and $(x,y)$ are oblate spheroidal coordinates related to the Weyl coordinates by
\[\rho^2=b^2 (x^2+1)(1-y^2), \quad z=bxy \,.\]
On the symmetry axis ($\rho=0$) where $x=\frac{|z|}{b}\,$, $y={\rm sign}\,z$,
$\frac{|y|}{\sqrt{x^{2}+1-y^{2}}}=\frac{1}{x}$ and $\frac{x}{\sqrt{x^{2}+1-y^{2}}}=1$, the disc potentials simplify to
\[\nu_{\rm iMM}^{(m)}(\rho=0)=
 -\frac{2^{2m+1}(m!)^2}{\pi}\frac{{\cal M}}{|z|}
  \sum_{n=0}^{m}C^{(m)}_{2n}\,{\rm i}
                Q_{2n}\!\!\left(\frac{{\rm i}b}{|z|}\right).\]

Our second external source, the Bach-Weyl ring (of mass ${\cal M}$ and Weyl radius $b$), generates potential
\begin{equation}  \label{nuBW,alt}
  \nu_{\rm BW}
  =-\frac{2{\cal M}K(k)}{\pi l_2} \,,
  \qquad
  l_{1,2}=\sqrt{(\rho\mp b)^2+z^2} \;,
\end{equation}
where $K(k)=\int_0^{\pi/2}\frac{{\rm d}\alpha}{\sqrt{1-k^2\sin^2\alpha}}$
is the complete elliptic integral of the 1st kind, with modulus and complementary modulus given by
\[k^2=1-\frac{(l_1)^2}{(l_2)^2}=\frac{4b\rho}{(l_2)^2}\;, \qquad
  k'^2=1-k^2=\frac{(l_1)^2}{(l_2)^2} \;.\]
Especially on the axis $\rho=0$, one has $k=0$, $K=\pi/2$, so
$\nu_{\rm BW}=-\frac{{\cal M}}{z^2+b^2}\,$.

Our main goal has been to analyse the behaviour of the geodesic flow in dependence on parameters of the system, namely on the relative mass and radius of the disc/ring and on energy and angular momentum of the particles.
It should be admitted that both sources are singular (the discs are 2D in space and the ring is even 1D) and curvature scalars really diverge at the ring as well as at the inner edge of the first iMM disc which we consider mostly (the higher-$m$ is the disc, the less singular it is at the edge). Hence, it is desirable to exclude from study the orbits which would approach the ring or the inner edge of the disc too closely, because there the space-time hardly corresponds to the astrophysical fields we want to approximate. Anyway, we assume that the test particles do not interact with the external source {\em mechanically} (in particular, they traverse the disc without collision).

\section{Geodesic chaos in the black-hole--disc/ring fields}
\label{chaos-methods}

Turning first to Poincar\'e surfaces, we were recording transitions of suitably chosen large sets of bound orbits through the equatorial plane (the plane of symmetry defined by the disc or ring) and drew the passages in terms of a radial component of four-velocity against radial coordinate. The results showed typical properties of a weakly non-integrable dynamical system, well described by the KAM theory. They were also confirmed by power spectra which we computed from vertical-coordinate time series, and by evolution of the ``latitudinal action" given by integral of the latitudinal component of momentum over an orbital period. (The character of dynamics can be well estimated from the time series itself, as well as from a spatial plot of the orbit.) Without repeating details from \cite{SemerakS-10}, let us list our main observations:
\begin{itemize}
\item
Geodesic flow tends to irregularity with increasing relative mass of the external source. The strongest chaos typically occurs when the disc mass is comparable to that of the black hole (for the ring it happens about 1/10 of the black-hole mass). For still heavier external source, the system rather returns to a more regular behaviour.
\item
Similarly with the dependence on particle energy: chaos first develops with ${\cal E}$ increasing, but for very high values it rather reduces back.
\item
The above are just overall tendencies, however the change of the system with parameters is by no means smooth. Indeed, one of the most typical and interesting aspects of chaotic dynamics is its {\em chaotic} dependence on parameters, with abrupt changes of the phase portrait occurring/disappearing within very narrow parameter ranges.
\item
The dependence on angular momentum $\ell$ is opposite: larger $\ell$ means larger part of energy allotted to azimuthal motion; and this component of motion is ``held" exactly by the conserved value of	$\ell$, so its larger value favours regularity.
\item
The ring being stronger source than the disc, it also induces stronger perturbation of the geodesic dynamics. In particular, the ring presence within the region accessible to the particles generates so many higher-periodic regular islands (surrounded by chaotic layers) that the resulting Poincar\'e diagrams can (also) be used as sophisticated wallpapers. However, if close encounters of the particles with the ring are prevented, the geodesic flow gets only moderately chaotic.
\item
On (equatorial) Poincar\'e sections, the geodesic dynamics rather tends to break up from the boundaries of the accessible phase-space region (these boundaries correspond to a zero vertical/latitudinal component of velocity, thus to a motion within the equatorial plane), while a certain regular region often survives in the interior.
\item
Chaos is a ``global" phenomenon and its notion is being used in connection with sufficient time spans, yet it is possible (and practically even inevitable) to speak of a degree of (ir)regularity of a certain restricted {\em section} of an orbit. Different sections of the same orbit may display very different degrees of chaoticity and the dynamics may switch between these modes quite suddenly. Orbits with such a variable behaviour are typical for weakly perturbed systems and at the same time they are most interesting for comparing different methods.
\item
Consistently with experience from the literature, such ``weakly chaotic" (parts of the) orbits which at times range over the chaotic ``sea" but also linger for a considerable intervals very close to regular regions of phase space (they ``stick" to them, hence {\em sticky} orbits) rather produce ``1/frequency" spectrum (i.e. approximated by a straight line in the log-log plot), whereas the ``strongly chaotic" (parts of the) orbits typically produce ``cat-back" spectral profile with (say, 100-times) less power at the low-frequency end, with less distinct peaks and more of tiny irregularities.
\end{itemize}

In \cite{SemerakS-12} we subjected the above dynamical system to two recurrence methods. The first of them, abbreviated WADV (from ``weighted average of directional vectors"), was proposed by \cite{KaplanG-92} as a way to distinguish between deterministic and random systems. Employing it within general relativity for the first time (as far as we know), we found it is even fairly sensitive to different degrees of (deterministic) chaos, especially in the regime of very weak perturbations (which we are mainly interested in). Briefly speaking, one takes a time series of some dynamical variable (we take vertical coordinate $z(\tau)$ as a function of the particle proper time $\tau$) and reconstructs the (3D) phase space from $z(\tau)$, $z(\tau-\Delta\tau)$, $z(\tau-2\Delta\tau)$, where $\Delta\tau$ is some time delay; this space is then ``rasterized" into a grid of boxes of some chosen size. The main point is to add (unit) directions in which orbits recurrently cross the prescribed boxes, average the length of the accumulated vector over the cells with respect to the number of passages and analyse its dependence on $\Delta\tau$ (and on the box size). Finally, the result is evaluated against its counter-part obtained for unit-step random walk, namely, one computes
\begin{equation}
  \bar{\Lambda}=
  {\rm average}\left\{\frac{\left(\frac{V_j}{n_j}\right)^2-\left(\bar{R}_{n_j}^d\right)^2}
                           {1-\left(\bar{R}_{n_j}^d\right)^2}\right\}
                     _{{\rm over~cells~(over~}j)}  \;,
\end{equation}
where
$V_j$ is the norm of the vector accumulated after $n_j$ passages through the $j$-th box and
\[\bar{R}^d_n
  =\frac{\Gamma\!\left(\frac{d+1}{2}\right)}{\Gamma\!\left(\frac{d}{2}\right)}
   \;\sqrt{\frac{2}{nd}} \;, \qquad
  {\rm in~particular} \quad
  \bar{R}^3_n=\frac{4}{\sqrt{6\pi n}}\]
denotes the average shift per step for random walk in $d$ dimensions, obtained for $n$ steps
(with particular value given for $d=3$ which is relevant in our case).
Now, the $\bar{\Lambda}$ realizes a normalized autocorrelation parameter: in theoretical limit (infinite evolution and infinitely fine grain), $\bar{\Lambda}=1$ for a deterministic signal, whereas $\bar{\Lambda}=0$ for random data; in practice, $\bar{\Lambda}(\Delta\tau)$ more or less decreases from 1, the faster the more chaotic (or even random) is the evolution.

The second recurrence method we applied rests on recurrences themselves of the orbits to chosen neighbourhoods of their past points. Denoting by ${\mathbf X}_{(i)}={\bf X}(\tau_{(i)})$ the $N$ successive points of a phase-space trajectory,\footnote
{The series of just one variable (e.g. position) suffices actually, since the phase space can be reconstructed from a sequence of its time-delayed copies as in the WADV method summarized above.}
the recurrences are simply recorded in the so called recurrence matrix
\begin{equation}
  R_{i,j}(\epsilon)=
  \Theta\left(\epsilon-\parallel\!\!{\bf X}_{(i)}-{\bf X}_{(j)}\!\!\parallel\right),
  \qquad i,j=1,...,N \;,
\end{equation}
where $\epsilon$ is the radius of a chosen neighbourhood (the selected threshold of ``close return"), $\parallel\cdot\parallel$ denotes the chosen norm (the picture of long-term dynamics only slightly depends on which one is used) and $\Theta$ is the Heaviside step function. The matrix thus contains only units and zeros and can be easily visualized by filling black dots (units) or blank spaces (zeros) at the respective coordinates $(i,j)$; such figures are called {\it recurrence plots} and were introduced by \cite{EckmannOR-87}. For regular systems, the recurrences (black points) arrange in distinct structures, in particular in long parallel diagonal lines (their distance scales with period), and weak chaos brings checkerboard structures, whereas for random systems the recurrences are scattered without order; chaotic (deterministic) systems yield the most interesting plots, consisting of rectangular blocks of almost-diagonal patterns as well as irregular ones, often looking like placed one over another.
The pattern of recurrences yields rich and credible data which can be further processed in numerous ways (see \cite{MarwanRTK-07} for a thorough survey). Within general relativity, the method has been employed e.g. by \cite{KopacekKKS-10,KovarKKK-13}.

Finally, one of the most characteristic symptoms of chaos is a sensitive dependence on initial conditions, following from the fast deviation of trajectories in the phase space. This divergence can be quantified by various indicators, of which Lyapunov characteristic exponents ($\lambda_i$) are the most well known, but other useful suggestions have also been given. The divergence quantifiers can be computed in two ways, either by following two nearby trajectories and the evolution of their phase-space separation, or by solving an appropriate variational equation (geodesic-deviation equation in our case) along one trajectory. We have adhered to the first, two-particle approach, using the procedure proposed --- within general relativity --- by \cite{WuH-03,WuHZ-06}. They argued, in particular, that it is sufficient to compute the orbital divergence in configuration space only, without including the momenta. This claim seems plausible and we have followed it, but a careful verification is still to be performed. Anyway, of the Lyapunov exponents, each characterizing the rate of separation change in a certain ($i$-th) independent direction, the most important is the maximal one (which prevails automatically in a longer evolution),
\begin{eqnarray}
  \lambda_{\rm max}=
  \lim_{\tau\to\infty} \frac{1}{\tau}\,
  \ln\frac{|\Delta\vec{x}(\tau)|}{|\Delta\vec{x}(0)|} \,, \qquad {\rm where} \quad
  |\Delta\vec{x}(\tau)|=\sqrt{|g_{\mu\nu}\Delta x^\mu\Delta x^\nu|(\tau)} \;.
\end{eqnarray}
We use small $x$ for the position in configuration space, ${\bf X}\!\equiv\!(x^\mu,p^\alpha)$, so $\Delta\vec{x}$ is the separation vector whose norm represents the momentary proper distance between the two neighbouring orbits. If $\lambda_{\rm max}>0$, than at least one direction exists in which the nearby orbits deviate exponentially.\footnote
{The exponents should reveal the nature of the flow {\em in the vicinity} of the reference world-line, hence, while time is running, the separation vector has to be renormalized whenever it reaches a certain ``too large" value; the velocity deviation vector, given by difference between four-velocities of the neighbouring world-lines, has to be renormalized by the same factor.}

Due to typically very slow convergence of the above limit, a number of related quantifiers has been proposed whose computation converges faster and which thus reveal the nature of orbits in a significantly shorter integration time. We have tried two of them, often used in celestial and galactic mechanics, the {\it fast Lyapunov indicator} (FLI) and the {\it mean exponential growth of nearby orbits} (MEGNO).
The FLI, suggested by \cite{FroeschleLG-97}, is given by
\begin{equation}
  {\rm FLI}(\tau)=\log_{10} \frac{|\Delta\vec{x}(\tau)|}{|\vec{x}(0)|}
\end{equation}
(restricting only to the configuration-space separation again).
FLI($\tau$) grows considerably faster for chaotic than for regular trajectories and this trend is evident much earlier than $\lambda_{\rm max}(\tau)$ approaches its limit value. In general relativity (motion in black-hole fields), the FLI has been employed by \cite{Han-08a,Han-08b}.
The second indicator, MEGNO, was proposed by \cite{CincottaS-00} as
\begin{equation}
  Y(T)=\frac{2}{T}\int\limits_0^T
       \frac{1}{|\Delta\vec{x}(\tau)|}\,
       \frac{{\rm d}|\Delta\vec{x}(\tau)|}{{\rm d}\tau}\;\tau\,{\rm d}\tau \;.
\end{equation}
The content of this quantity is simply its {\em value} (rather than rate of growth or even ``irregularity of behaviour"), which makes it suitable for automatic surveys over large areas of phase space. Importantly, the MEGNO distinguishes between regular and chaotic evolutions securely, because for regular orbits it tends to 2 (with an additional bounded oscillating term), whereas for chaotic orbits it grows linearly, with a slope corresponding to the value of the maximal Lyapunov exponent ($Y\approx\lambda_{\rm max}T$ for large enough proper time). A few years ago, \cite{MestreCG-11} found an analytic relation between FLI and MEGNO,
\begin{equation}
  Y(T) = 2\,[{\rm FLI}(T)-\overline{\rm FLI}(T)]\,\ln(10) \,,
\end{equation}
where $\overline{\rm FLI}(T)$ is the FLI time averaged over the period $\langle 0,T\rangle$,
\[\overline{\rm FLI}(T)=\frac{1}{T}\int\limits_0^T {\rm FLI}(\tau)\,{\rm d}\tau \,.\]
Note that one can in turn average the MEGNO over some period in a similar manner (cf. the last part of figure \ref{fig08}).

We presented the results obtained using Lyapunov exponents, FLI and MEGNO in paper \cite{SukovaS-13}, together with more details and remarks e.g. on often queried (non-)invariance of the world-line deviation indicators. A thorough comparison of these quantities with other similar indicators has recently been given by \cite{Lukes-G-14} using the variational approach.

\subsection{Relations between chaotic indicators --- an example}

Besides relations like that between the FLI and MEGNO indicators (which are of similar nature), yet more interesting are relations between quantities whose origins are more independent. One illustration is an estimate of the maximal Lyapunov exponent which can be obtained from the MEGNO slope as indicated above, but also (for instance) from the 2nd-order R\'enyi's entropy (also called correlation entropy), one of the indicators derived from the recurrence analysis. Namely, \cite{MarwanRTK-07} showed that this quantity yields a lower estimate of the sum of positive Lyapunov exponents, and they also suggested that it can be approximated by
\begin{equation}
  K_2(\epsilon,l)\approx -\frac{1}{l\Delta\tau}\ln{p_c(\epsilon,l)}\,,
\end{equation}
where $\Delta\tau$ is the sampling time step and $p_c(\epsilon,l)$ is the probability of finding a diagonal line whose length is at least $l$ in the recurrence matrix ($\epsilon$ is the radius of a chosen ``close neighbourhood"). This means that $K_2$ is estimated from a slope of the cumulative histogram of diagonal lines (of diagonals at least $l$ points long), plotted in logarithmic scale against the length $l$. We computed both estimates (from MEGNO slope and from R\'enyi's entropy) and confirmed that they are in line with the values of $\lambda_{\rm max}$ obtained by direct computation.

Let us make a remark concerning the ``value/price ratio" of the above indicators. Whereas MEGNO was introduced as a handy ``condensate" of Lyapunov exponents which can be quite easily implemented in automated scans, the R\'enyi's entropy is more sophisticated, but also less accessible for routine computer evaluation, mainly because it has to be determined from a proper part of the histogram. Namely, only a certain middle section of the histogram, reasonably following a straight line, is relevant, since the short-length end typically diverges due to increasing number of ``sojourn" points (successive points between which the orbit does not leave the given $\epsilon$-neighbourhood\footnote
{These points does {\em not} represent true recurrences and are usually discarded from the statistics.}),
while the long-length end typically falls off quickly due to a finite length of the trajectory. One should stress that the histogram is theoretically to be computed in the limit $l\rightarrow\infty$, so the short-length end has actually no sense, while the long-length end is of course determined by the fact that practically the trajectories cannot be infinitely long.
We are depicting this to point out the benefits of some simpler recurrence quantifiers, especially of the one called DIV, given by inverse of the length of the longest diagonal. We compared the $K_2$ entropy with the DIV, in particular, we coloured several Poincar\'e diagrams according to the values of $K_2$ and DIV obtained for the orbits followed in a given run, and observed that both the quantifiers provide virtually the same information (cf. figure \ref{fig07}).

\section{Numerical illustrations}

\begin{figure*}
\includegraphics[width=\textwidth]{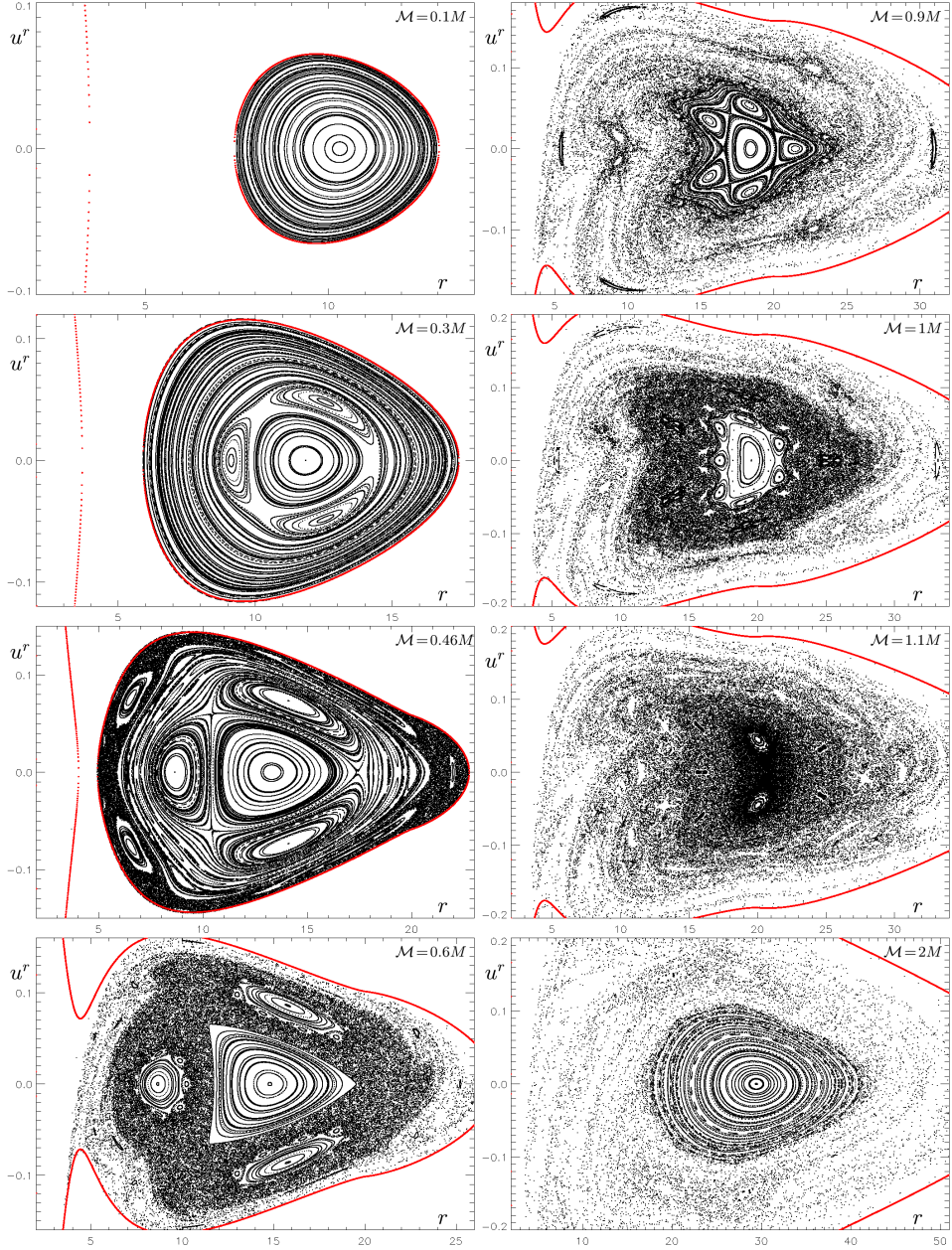}
\caption
{Passages of geodesics with $\ell\!=\!3.75M$ and ${\cal E}\!=\!0.955$ through the equatorial plane of a Schwarzschild black hole (with mass $M$) surrounded by the inverted 1st Morgan-Morgan disc (with inner Schwarzschild radius $r_{\rm disc}\!=\!20M$), drawn for different relative disc masses ${\cal M}/M$ (indicated in the plots). Red line bounds the accessible region. Adopted from \cite{SemerakS-10}.}
\label{fig01}
\end{figure*}

\begin{figure*}
\includegraphics[width=\textwidth]{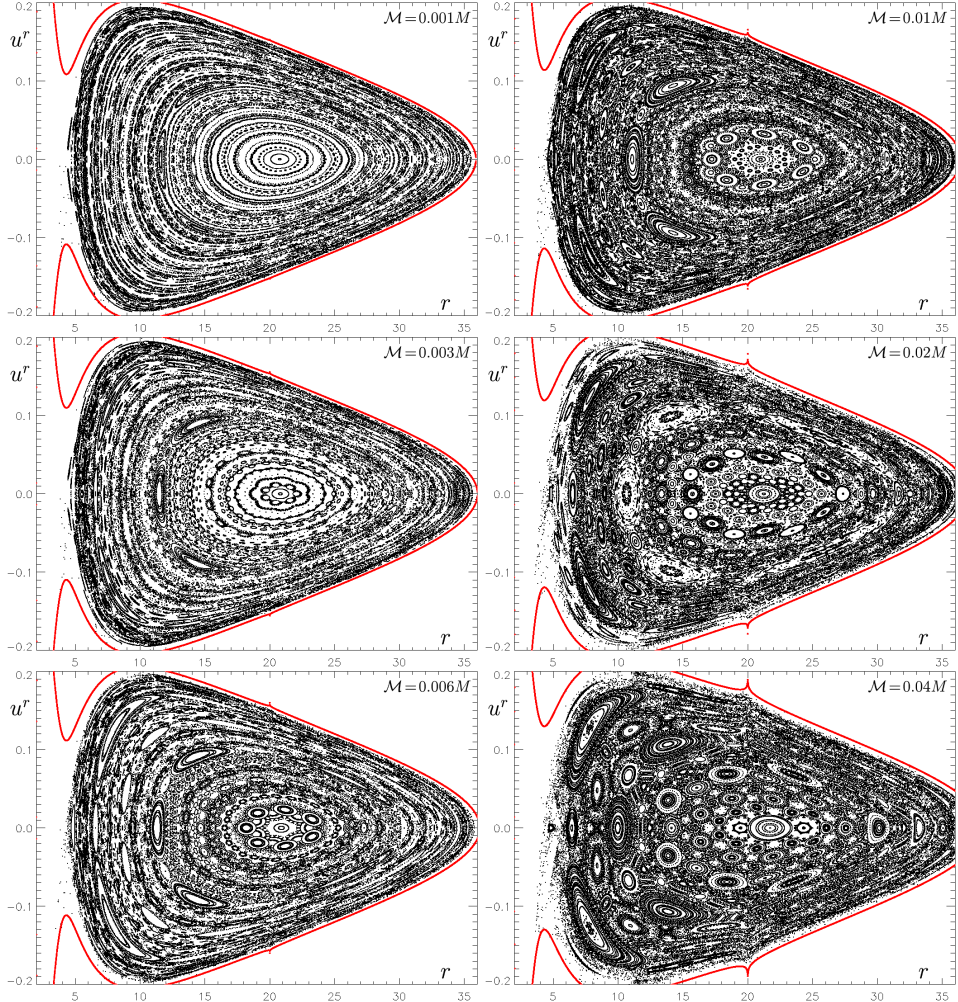}
\caption
{Passages of geodesics with $\ell\!=\!3.75M$ and ${\cal E}\!=\!0.977$ through the equatorial plane of a Schwarzschild black hole (with mass $M$) surrounded by a Bach-Weyl ring (with Schwarzschild radius $r_{\rm ring}\!=\!20M$), drawn for different relative ring masses ${\cal M}/M$ (indicated in the plots). Red line bounds the accessible region. The figure continues on the next page.}
\end{figure*}

\begin{figure*}
\includegraphics[width=\textwidth]{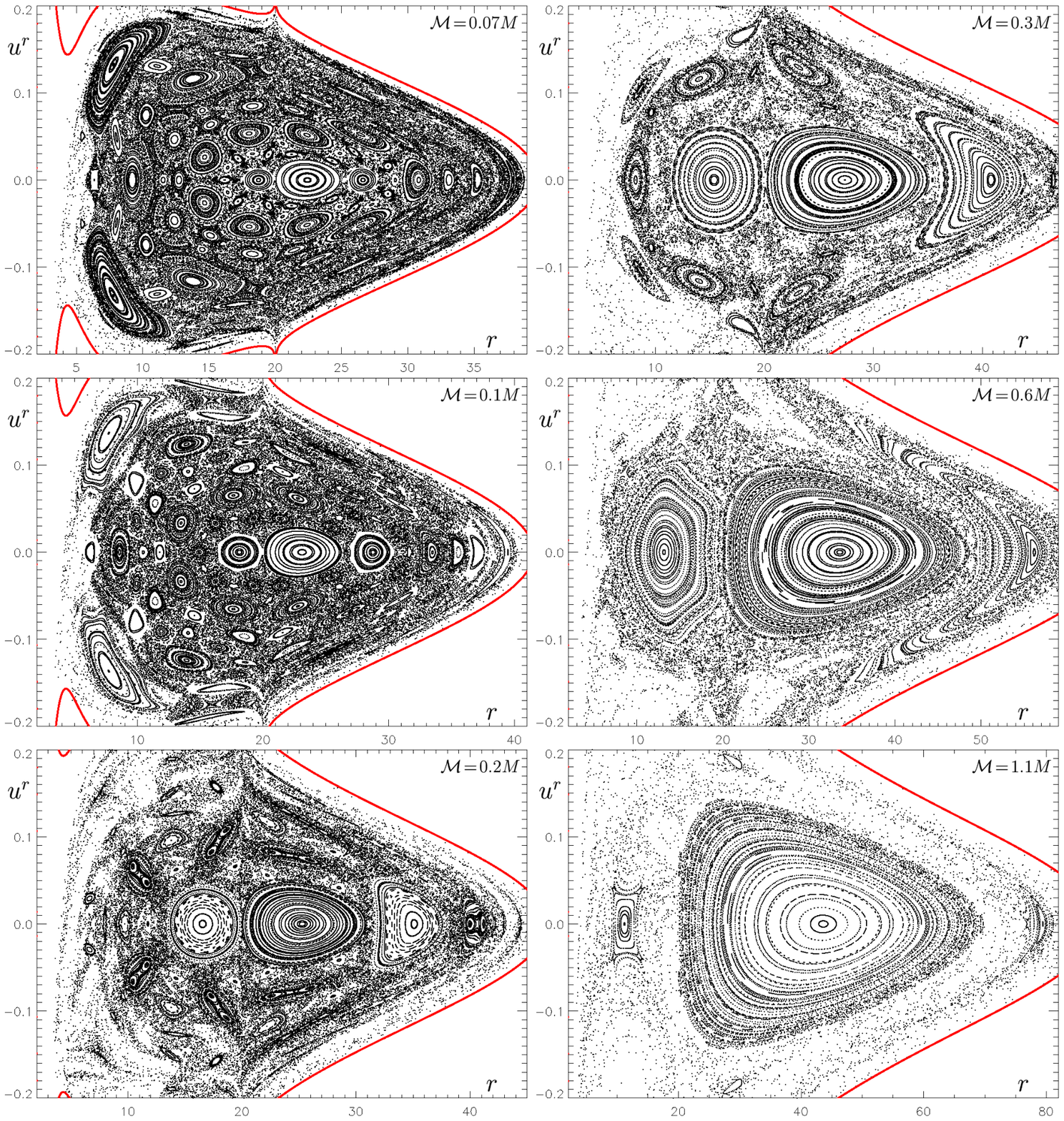}
\ContinuedFloat
\caption
{{\it (Continued:)} With growing mass of the ring, the phase space apparently becomes very complex, containing many islands formed by higher-periodic regular orbits, interwoven with chaotic layers. For very high relative masses, the primary regular island restores and dominates the plot again. However, if the particles could not get close to the ring (if the latter did not lie within their accessible region), their motion would be considerably less irregular. The figure is adopted from \cite{SemerakS-10}.}
\label{fig02}
\end{figure*}

\begin{figure*}
\includegraphics[width=0.862\textwidth]{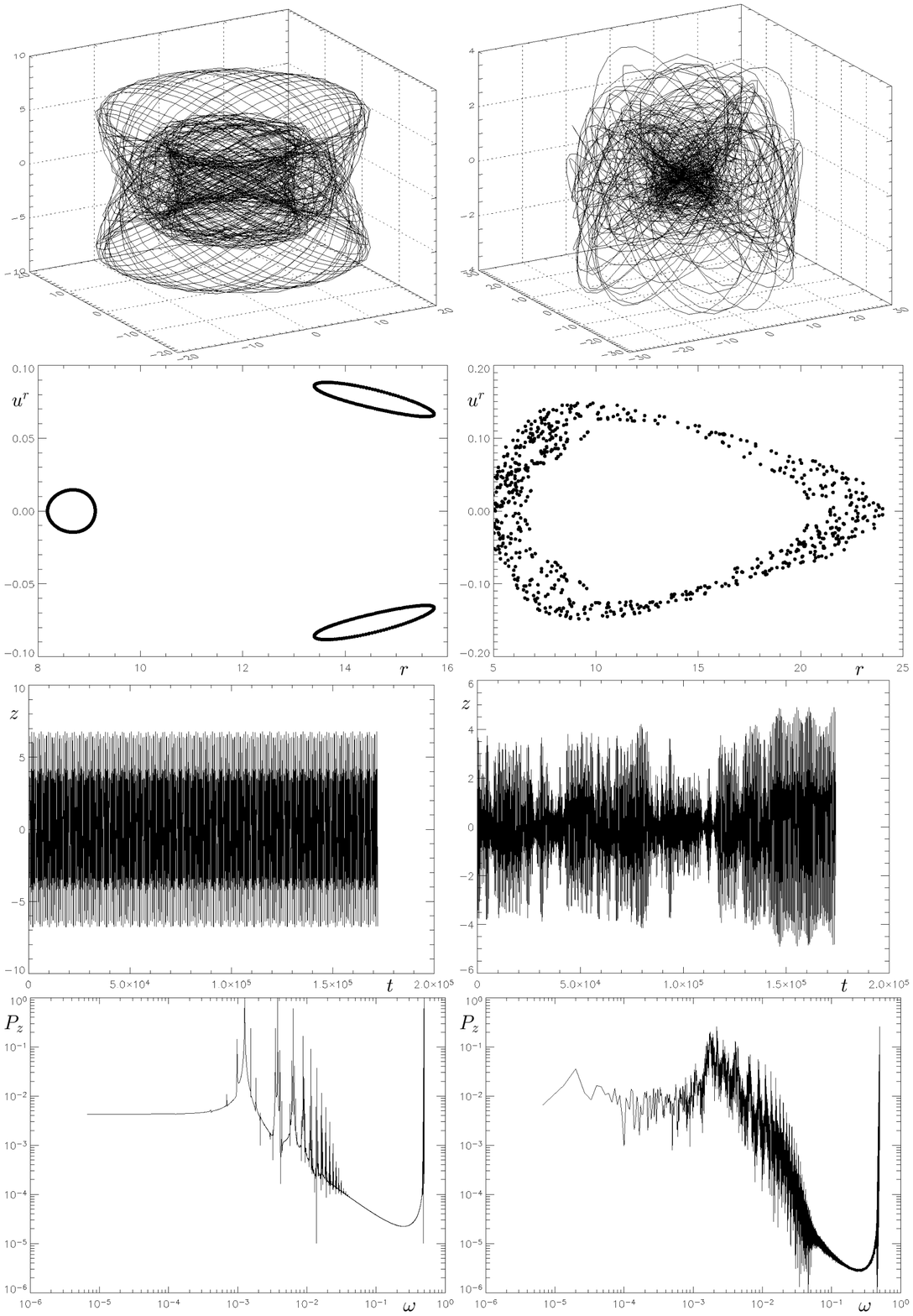}
\caption
{Example of a regular ({\it left}) and chaotic ({\it right}) geodesic, both with ${\cal E}\!=\!0.955$ and $\ell\!=\!3.75M$, in the field of a Schwarzschild black hole surrounded by the inverted 1st Morgan-Morgan disc with ${\cal M}\!=\!0.5M$, $r_{\rm disc}\!=\!20M$. From top to bottom, the rows show their spatial tracks, Poincar\'e sections $z\!=\!0$ ($r$,$u^r$), time series of the $z$ position and corresponding power spectra. Coordinates are in $[M]$, frequencies in $[1/M]$. Adopted from \cite{SemerakS-10}.}
\label{fig03}
\end{figure*}

\begin{figure*}
\includegraphics[width=\textwidth]{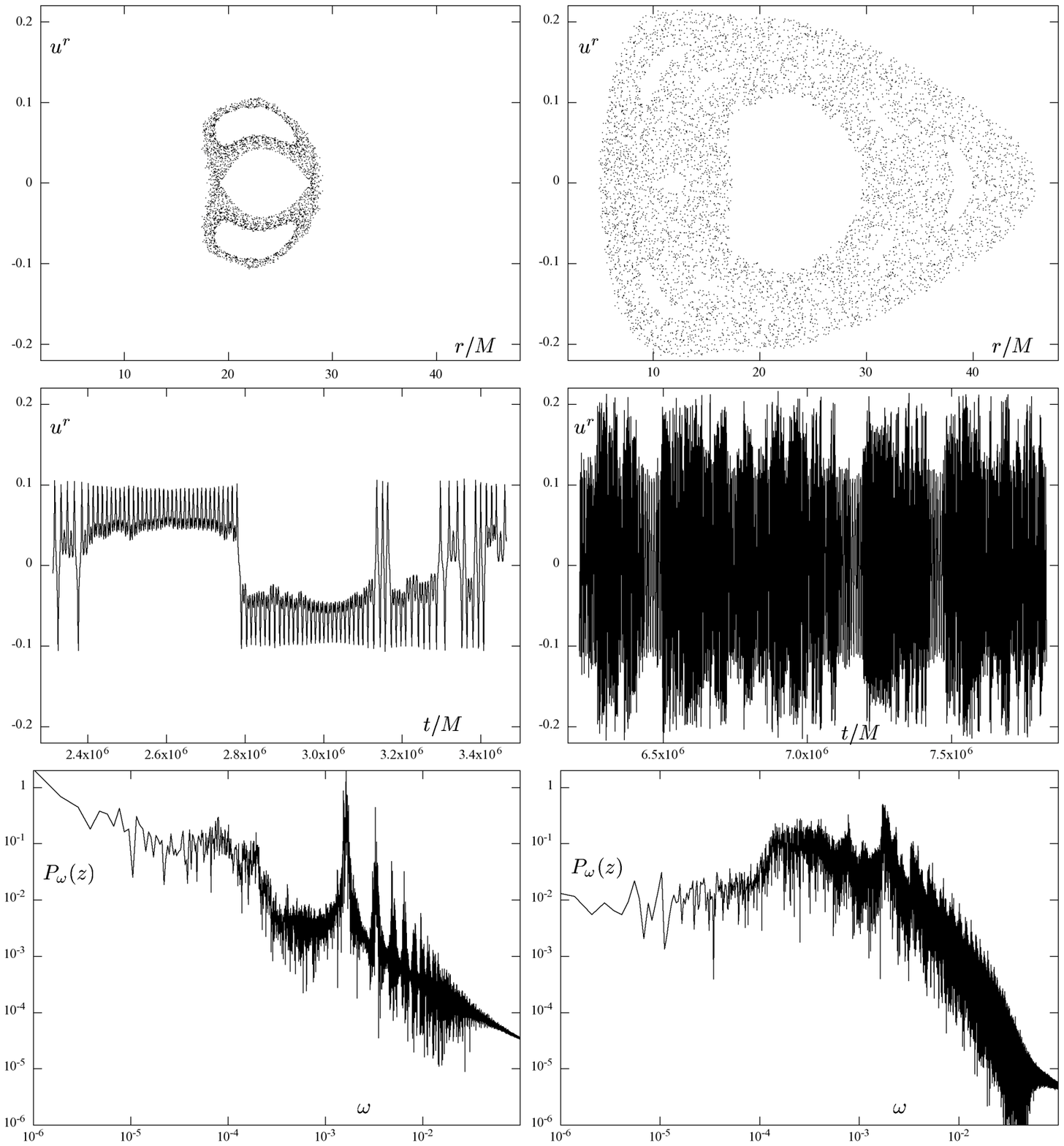}
\caption
{Example of a ``sticky" (left column) and strongly chaotic (right column) phases of one and the same geodesic (having ${\cal E}\!=\!0.956$ and $\ell\!=\!4M$) in the field of a black hole plus an inverted 1st Morgan-Morgan disc (with ${\cal M}\!=\!1.3M$, $r_{\rm disc}\!=\!20M$): Poincar\'e sections $z\!=\!0$ ($r$,$u^r$), time series of $u^r$ and power spectra of the $z$ position. Adopted from \cite{SemerakS-12}.}
\label{fig04}
\end{figure*}

\begin{figure*}
\includegraphics[width=\textwidth]{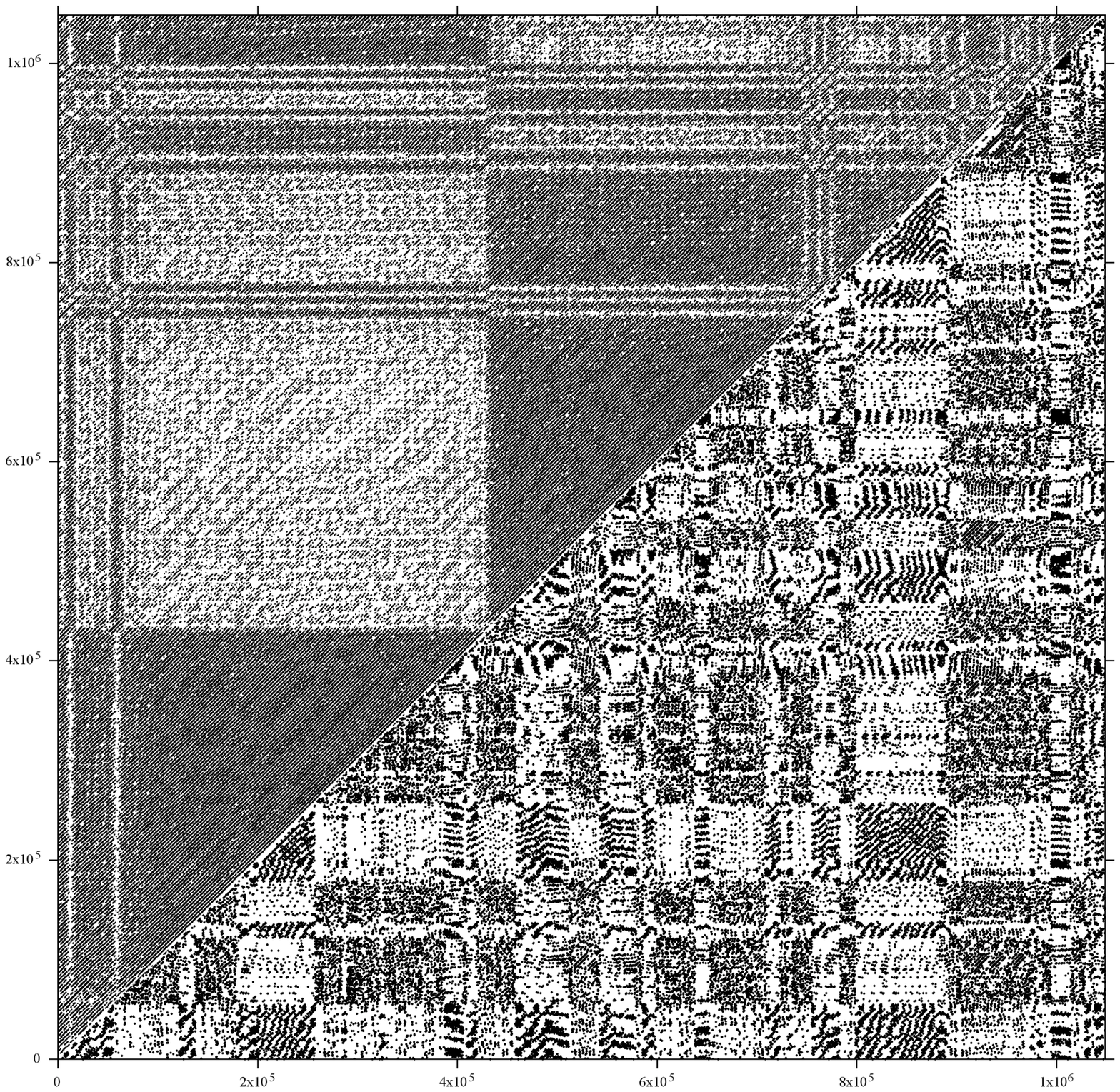}
\ContinuedFloat
\caption
{{\it (Continued:)} Recurrence plots for the same two orbital sections. Considering the recurrence-matrix symmetry, we give just halves of the plots in one square: the weakly chaotic section is above the main diagonal, while the strongly chaotic one is below the main diagonal. The axis values indicate proper time in units of $M$. Adopted from \cite{SemerakS-12}.}
\label{orbit1-excerpts}
\end{figure*}

\begin{figure*}
\includegraphics[width=\textwidth]{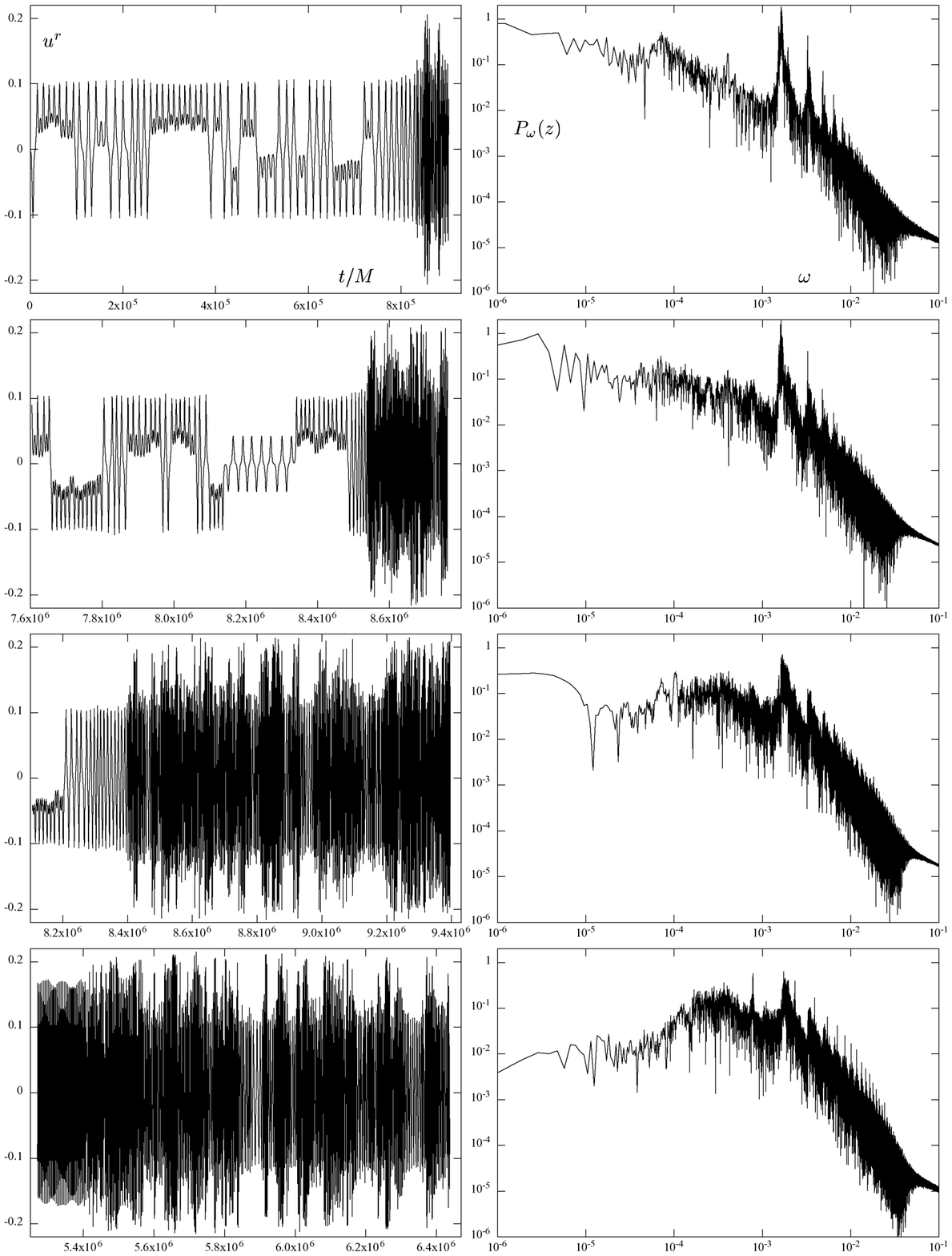}
\caption
{The $u^r(t)$ behaviour (left column) and power spectra of $z(t)$ evolution (right column) of four different sections of the same orbit as in figure \ref{orbit1-excerpts}. Chaoticity clearly grows from top to bottom. Adopted from \cite{SemerakS-12}.}
\label{fig05}
\end{figure*}

\begin{figure*}
\includegraphics[width=\textwidth]{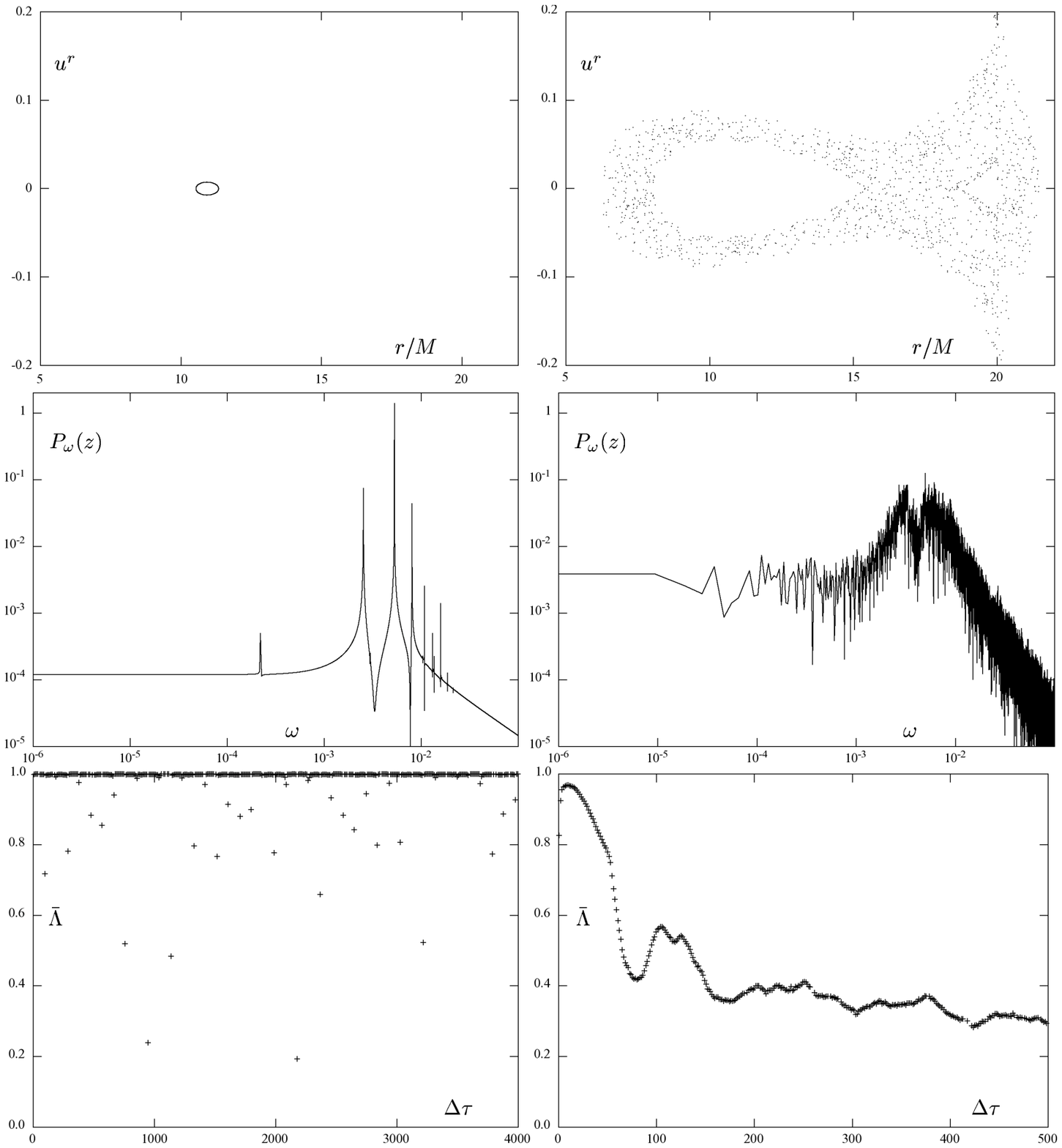}
\caption
{Difference between a geodesic lying deep in a primary regular island (left column) and the one living in a chaotic sea (right column), as seen on Poincar\'e diagram (top), $z(t)$ evolution (middle) and the Kaplan-Glass averaged autocorrelation parameter $\bar\Lambda(\Delta\tau)$ (bottom).
Both geodesics have ${\cal E}\!=\!0.93$ and $\ell\!=\!3.75M$ and the background is determined by a black hole surrounded by a Bach-Weyl ring with ${\cal M}\!=\!0.5M$ and $r_{\rm ring}\!=\!20M$. Adopted from \cite{SemerakS-12}.}
\label{fig06}
\end{figure*}

\begin{figure*}
\includegraphics[width=\textwidth]{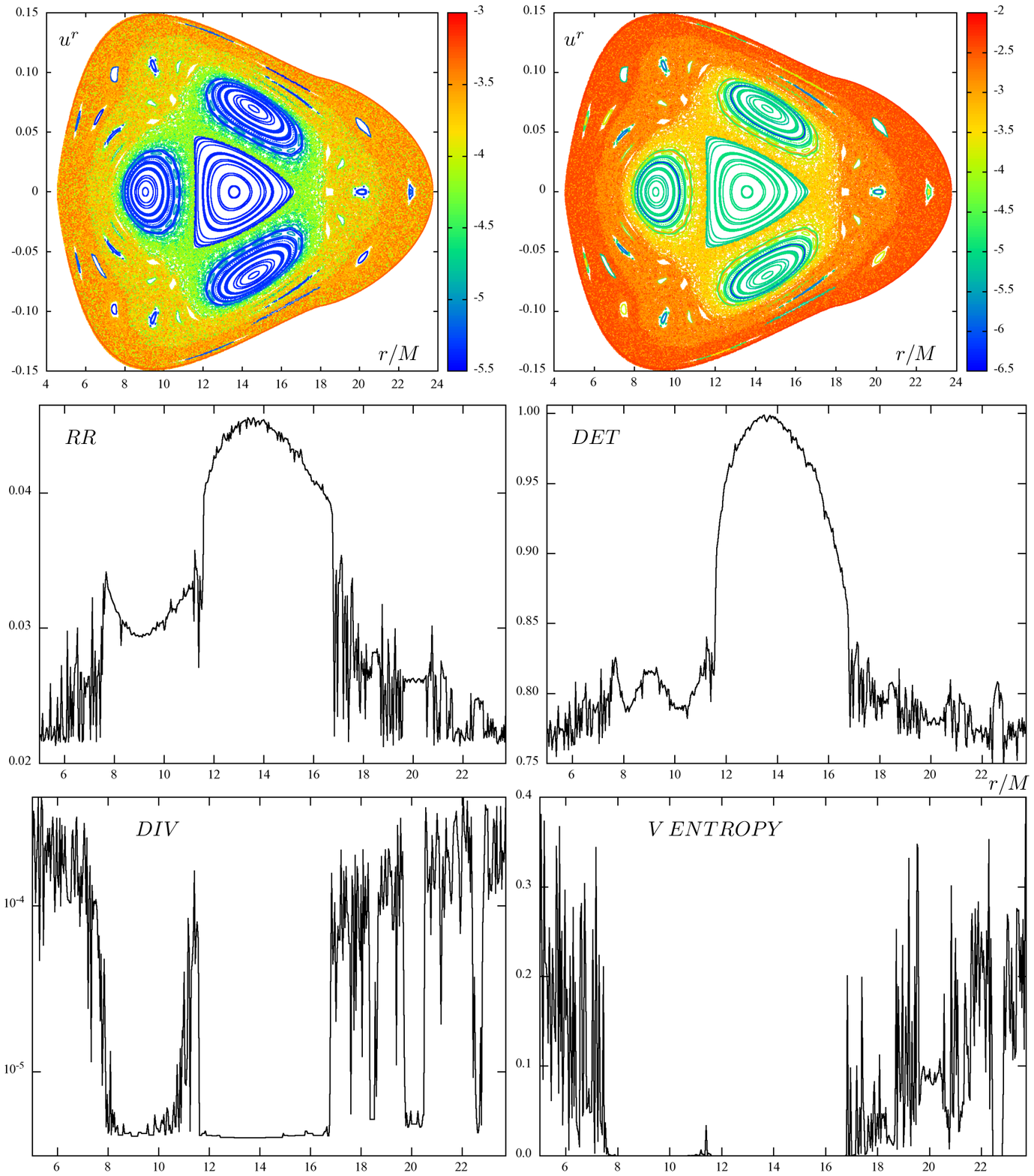}
\caption
{Examples of quantifiers extracted from the recurrence matrix, as computed for 470 geodesics launched with ${\cal E}\!=\!0.9532$, $\ell\!=\!3.75M$ tangentially (with $u^r\!=\!0$) from radii between $5M$ and $24M$ from equatorial plane of a black hole ($M$) surrounded by the inverted 1st MM disc (${\cal M}\!=\!0.5M$, $r_{\rm disc}\!=\!18M$). The orbits were followed for about $250000M$ of proper time with ``sampling" $\Delta\tau\!=\!45M$. In Poincar\'e diagram (top), they are coloured according to their DIV (left) and $K_2$ (right) values (log scale).
The quantifiers shown (see text for description) are clearly sensitive to tiny phase-space features (behaviour along the top-diagram $r$-axis).
Adopted from \cite{SemerakS-12}.}
\label{fig07}
\end{figure*}

\begin{figure*}
\includegraphics[width=\textwidth]{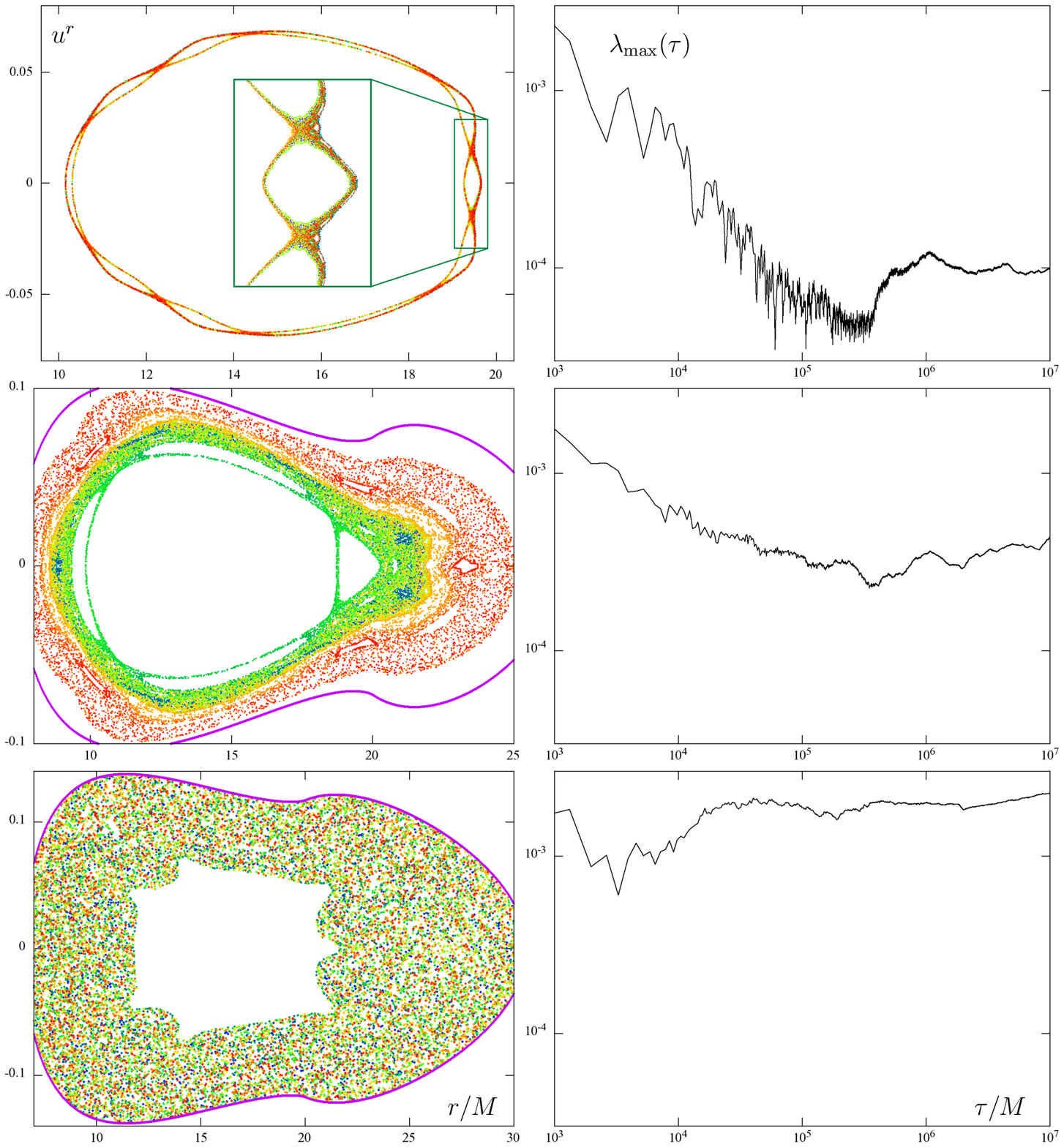}
\caption
{Poincar\'e sections (left) and maximal Lyapunov exponents (right) for three orbits (D, E, F) of different degrees of irregularity, with parameters ${\cal M}\!=\!0.94M$, $r_{\rm disc}\!=\!20M$, ${\cal E}\!=\!0.947$, $\ell\!=\!4M$ in the upper row (D); ${\cal M}\!=\!1.3M$, $r_{\rm disc}\!=\!20M$, ${\cal E}\!=\!0.9365$, $\ell\!=\!4M$ in the middle row (E); and ${\cal M}\!=\!1.3M$, $r_{\rm disc}\!=\!20M$, ${\cal E}\!=\!0.941$, $\ell\!=\!4M$ in the bottom row (F). The orbits fill phase-space layers of different volumes, in agreement with the obtained values $\lambda_{\rm max}(\tau_{\rm max})\!\doteq\!9.88\cdot 10^{-5} M^{-1}$ (top), $4.28\cdot 10^{-4}M^{-1}$ (middle) and $2.25\cdot 10^{-3}M^{-1}$ (bottom). The Poincar\'e-surface passages are coloured by proper time (it increases in the order blue $\rightarrow$ green $\rightarrow$ yellow $\rightarrow$ red).}
\label{fig08}
\end{figure*}

\begin{figure*}
\includegraphics[width=\textwidth]{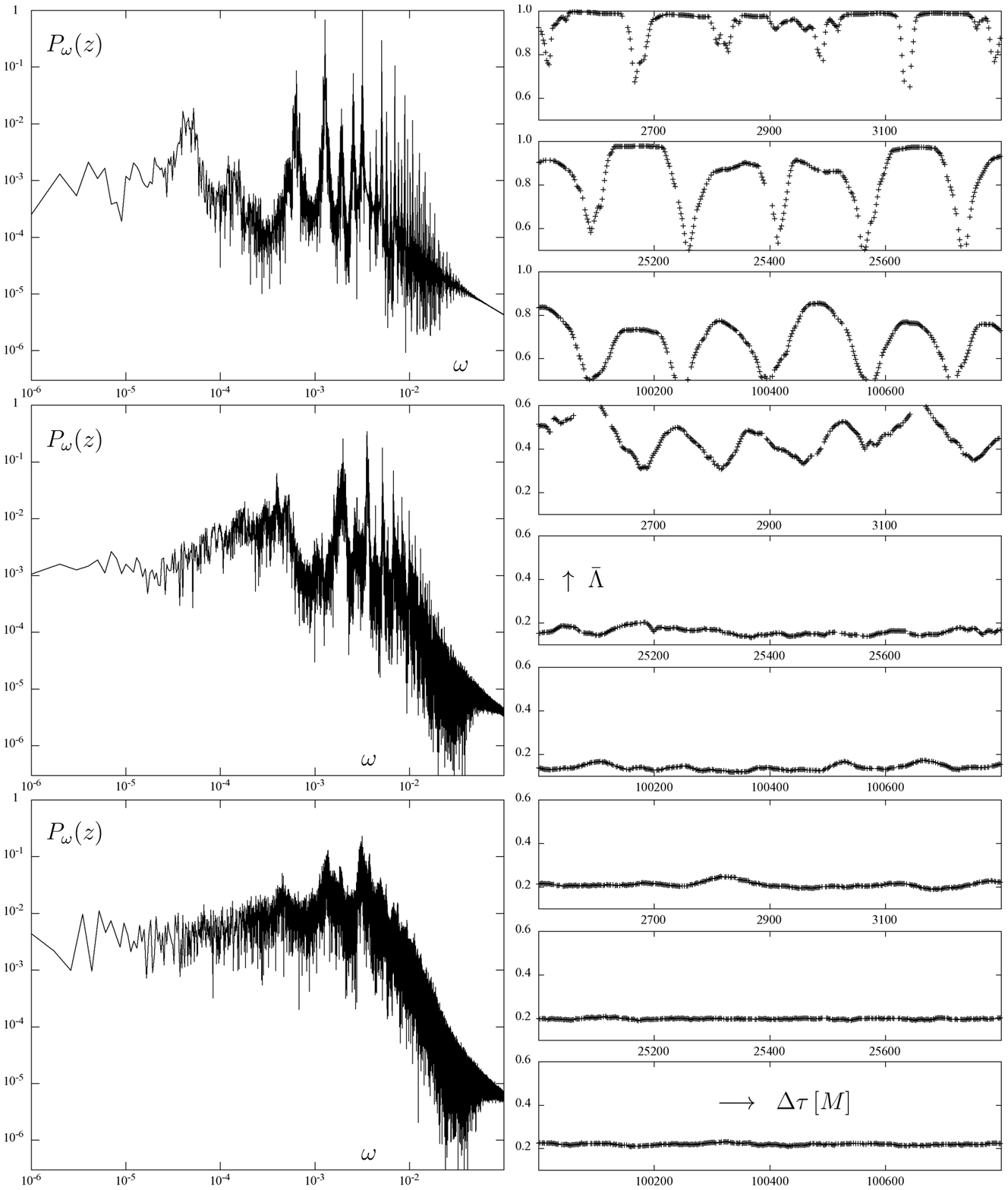}
\ContinuedFloat
\caption
{{\it (Continued:)} Power spectra of $z(t)$ (left column) and the Kaplan-Glass average of the recursion directions $\bar\Lambda$ (right column), plotted for the same three orbits D, E, F (for their parts up to $\tau_{\rm max}\!=\!10^6 M$). The $\bar\Lambda(\Delta\tau)$ dependence is drawn for three separate time-shift ($\Delta\tau$) intervals, (2500--3300)$M$, (25000--25800)$M$ and (100\,000--100\,800)$M$. Note that vertical-axes ranges are different on the right, 0.5--1.0 for the top orbit whereas 0.1--0.6 for the remaining two.}
\end{figure*}

\begin{figure*}
\includegraphics[width=\textwidth]{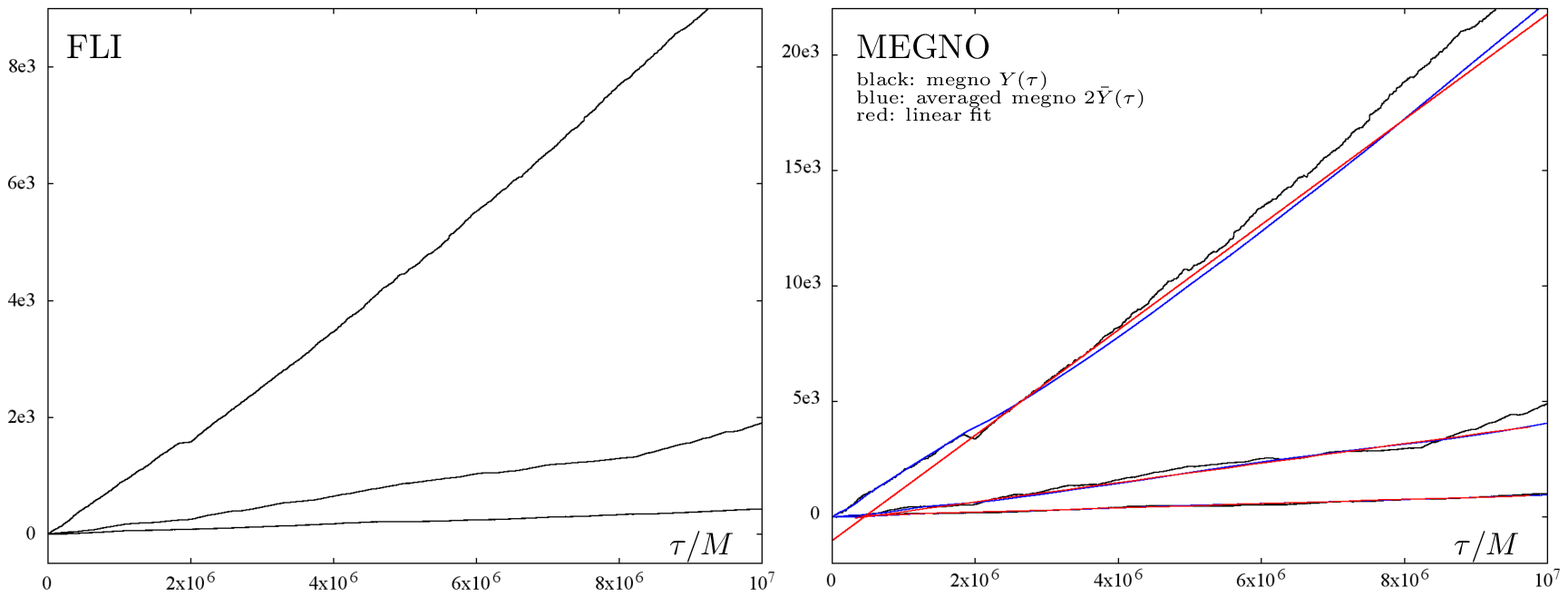}
\ContinuedFloat
\caption
{{\it (Continued:)} FLI($\tau$) on the left and $Y(\tau)$ (MEGNO) on the right, computed for the same three orbits (D, E, F). The average MEGNO $2\overline{Y}(\tau)$ is drawn in blue and its linear fit is in red. The maximal-Lyapunov-exponent values inferred from the average-MEGNO slope are $9.172\cdot 10^{-5}M^{-1}$ for orbit D (bottom), $4.192\cdot 10^{-4} M^{-1}$ for orbit E (middle) and $2.278\cdot 10^{-3}M^{-1}$ for orbit F (top).
All three plots are adopted from \cite{SukovaS-13}.}
\end{figure*}

\begin{figure*}
\includegraphics[width=\textwidth]{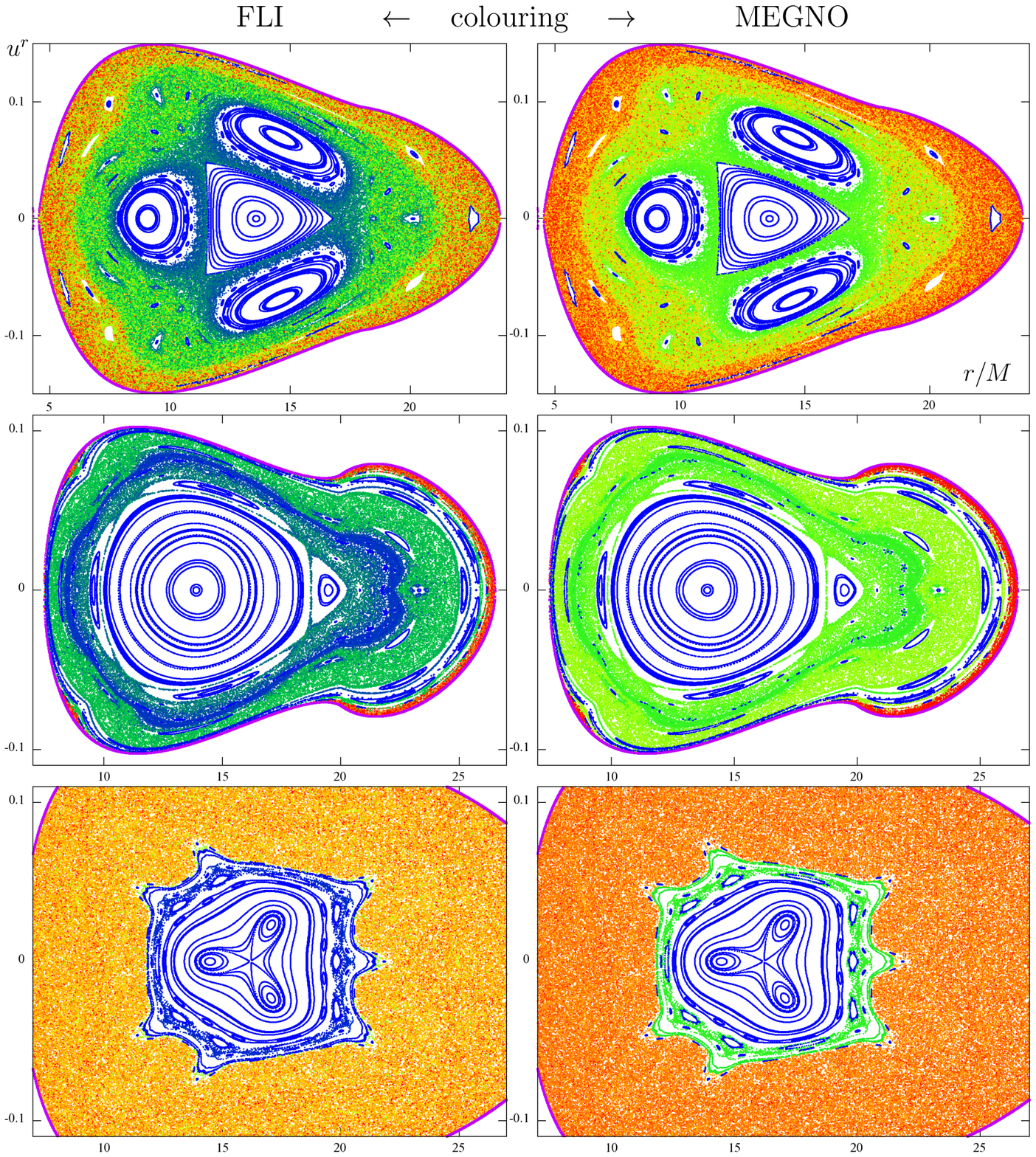}
\caption
{Equatorial Poincar\'e maps of three phase portraits, coloured according to FLI($\tau_{\rm max}$) (left column) and $\bar{Y}(\tau_{\rm max})$ (right column) values, where $\tau_{\rm max}\!=\!250\,000 M$. Parameters: ${\cal M}\!=\!M/2$, $r_{\rm disc}\!=\!18M$, ${\cal E}\!=\!0.9532$, $\ell\!=\!3.75M$ in the 1st row; ${\cal M}\!=\!1.3M$, $r_{\rm disc}\!=\!20M$, ${\cal E}\!=\!0.9365$, $\ell\!=\!4M$ in the 2nd row; and ${\cal M}\!=\!1.3M$, $r_{\rm disc}\!=\!20M$, ${\cal E}\!=\!0.941$, $\ell\!=\!4M$ in the 3rd row. Whenever $\bar{Y}(\tau_{\rm max})\!>\!4$, we add another 200 to it in order to enhance distinction between regular and weakly chaotic regions. The colours going from blue to red in the visible-spectrum order correspond to FLI increasing from 0 to 350 (left) and to MEGNO increasing from 0 to 500. Adopted from \cite{SukovaS-13}.}
\label{fig09}
\end{figure*}

Let us give several examples of how the varying degrees of chaoticity are revealed by the methods we have mentioned above. The passages through a suitably chosen plane, plotted in a suitably chosen variables (called Poincar\'e maps), provide a direct, reliable and indicative picture of a studied motion. We naturally start with them, showing how the nature of geodesic flow around a black hole changes when a more and more massive thin disc (figure \ref{fig01}) or ring (figure \ref{fig02}) are placed around in a concentric manner. In both cases, the passages are recorded of some bunch of bound time-like geodesics through the equatorial plane of the system, and plotted in terms of the Schwarzschild-type  radius $r$ and of the corresponding component of four-velocity $u^r$.
Figure \ref{fig03} compares a regular geodesic with a chaotic one, using four representations: their spatial tracks, equatorial Poincar\'e sections, time dependence of the ``vertical" ($z$) position and the corresponding power spectra.
Figure \ref{fig04} compares two different parts of the {\em same} geodesic, one ``sticking" to a regular island and the other filling a chaotic sea. Four plots are shown --- equatorial Poincar\'e sections, time series of the $u^r$ four-velocity component, the power spectra of $z(t)$ and, in the second part of the figure, recurrence plots; the difference between the two orbital phases is clearly seen in all of them. Four different parts of the same orbit are also shown in figure \ref{fig05} and their increasing chaoticity revealed on the $u^r(t)$ evolution and on the power spectra of $z(t)$.

The above representations are illustrative for a human eye, but less suitable for a quantitative judgement about the ``degree of chaoticity" of the system, mainly if such a judgement should be entrusted to a computer. The recurrence methods we tested offer several ``quantifiers" that are more suitable in this respect and which assign to every single orbit a single number or a single dependence on some parameter. In the first method, the character of the orbit is recognized from the average of directions in which the orbit recurrently crosses the prescribed phase-space cells (the Kaplan-Glass WADV method described in previous section). In figure \ref{fig06}, this averaged autocorrelation parameter is compared for a geodesic lying deep in a primary regular island and for a one inhabiting a large chaotic region; also added are the respective Poincar\'e diagrams and $z(t)$ evolutions.
Figure \ref{fig07} presents several quantifiers computed from the recurrence matrix and requires some commentary. The quantifiers shown are called RR, DET, DIV and V$\,$ENTROPY (see \cite{MarwanRTK-07}). The {\it recurrence rate} RR is given by ratio of the recurrence points (black ones) within all points of the recurrence matrix,
\[{\rm RR}(\epsilon)=\frac{1}{N^2}\sum_{i,j=1}^N R_{i,j}(\epsilon) \,.\]
The DIV (``divergence") quantifier is just reciprocal of the length of the recurrence-matrix longest diagonal.
The DET (``determinism") is a ratio of the points which form a diagonal line longer than a certain $l_{\rm min}$ within all recurrence points,
\[{\rm DET}(\epsilon)=
  \frac{\sum_{l=l_{\rm min}}^N lP(\epsilon,l)}{\sum_{l=1}^N lP(\epsilon,l)} \;,\]
where $P(\epsilon,l)$ denotes histogram of lengths of the diagonal lines (length spectrum of the diagonals).
Finally, the V$\,$ENTROPY represents the Shannon entropy of the probability $p(v)$ that a vertical line has length $v$,
\[{\rm V\,ENTROPY} = -\!\!\sum_{v=v_{\rm min}}^N p(\epsilon,v)\ln p(\epsilon,v) \,, \qquad
  p(\epsilon,v)=\frac{P(\epsilon,v)}{\sum_{v=v_{\rm min}}^N P(\epsilon,v)} \,.\]
In figure \ref{fig07} we went along the $u^r=0$ axis of the Poincar\'e diagrams and plotted the value of the above quantifiers for geodesics starting tangentially from the respective radii; it represents 470 geodesics in total, starting from radii between $5M$ and $24M$. As already noted at the end of previous section, the colouring of the Poincar\'e maps by the values of the ``rough" quantity DIV and its ``sophisticated" relative $K_2$ shows that both bring practically the same information. It is also obvious that the quantifiers (including DIV) are quite sensitive and can uncover even tiny features of phase space.

In the last two figures, the quantifiers of geodesic deviation are illustrated and their message compared with that provided by previous methods. Three orbits of different degrees of chaoticity are analyzed in figure \ref{fig08}, first on Poincar\'e sections and by computing the maximal Lyapunov exponents, then on power spectra of $z(t)$ and on behaviour of the Kaplan-Glass directional autocorrelation in dependence on time shift applied to $z(t)$, and finally by computing the FLI and MEGNO indicators. All the methods clearly distinguish between the orbits and yield consistent results. Figure \ref{fig09} presents Poincar\'e maps of three different phase-space situations, coloured by FLI($\tau_{\rm max}$) and MEGNO($\tau_{\rm max}$). The quantities apparently provide almost the same message, but MEGNO is somewhat more helpful, because, due to its definite value of 2 for regular orbits, it is more precise in distinguishing them from the chaotic ones (whose MEGNO value is bigger); to benefit from this graphically, one can, for example, increase all the values of MEGNO indicating chaotic evolution by some constant (we did it with all values above 4, in order to be on the safe side).

\section{Concluding remarks}

In the Introduction, we indicated black-hole dominated cores of galactic nuclei as a natural astrophysical motivation. Indeed, the individual stars there can be approximated as point test particles and if the environment is sufficiently sparse, their motion is close to a geodesic one. In our Galaxy, for instance, there is a black hole of mass $M\doteq 4.3\cdot 10^6\,M_\odot$, probably surrounded by a rather tiny accretion structure and by {\em two} much larger circum-nuclear rings, one with radius 2 parsecs and mass $M/10$ and the other around the 80 parsec radius with mass $10M$. (These are average values learned from literature, see \cite{SukovaS-13} for references.) There is also a considerable nuclear star cluster which could be incorporated by adding a suitable spherical term in our potential $\nu$. Of the above structures, only the smaller circum-nuclear ring has been observed to be able to partially destabilize the motion of stars, and only if their orbits can get sufficiently close to it.

The dynamics of astrophysical systems clearly has observable consequences conveyed by electromagnetic as well as gravitational radiation. The character of dynamics is crucial for the long-term evolution of these systems, although examples like the motion of stars in galactic nuclei are happening over too long intervals to provide well detectable signatures of chaos in ``real time". The characteristic periods are of course the shorter the closer to the horizon the motion takes place and \cite{Lukes-GAC-10} argued that it could be possible to recognize whether the central object is Kerr-like or different from the radiation of an inspiralling captured geodesic orbiter. Without doubt, the astrophysical black holes {\em should} differ from the Kerr ideal for other reasons (than considered here) as well: they must be interacting (if only to be ``observable"), so not only they are not isolated, but also not stationary; and probably they do not live in an asymptotically flat universe. In any case, the study of motion in deformed black-hole fields can lead to fundamental questions concerning the nature of objects supposed to play a key role in the most engaging cosmic systems.

Let us conclude by several suggestions of further possible work.
One could certainly subject the system to still another methods and codes in order to check the results, but also to evaluate the methods/codes themselves and their practical features within general relativity; such analyses like \cite{Lukes-G-14} (on the usage of several indicators of orbital deviation in GR) will be helpful. For example, the Melnikov-integral method has already been employed several times in general relativity, as well as the study of the dimension of ``basin boundaries" (boundaries between initial-condition sets which evolve to distinct end states), which are of particular appeal due to their coordinate-independent message. It would be interesting to perform a similar analysis for the ``corresponding" (pseudo-)Newtonian or post-Newtonian systems, at least to infer how good approximations they provide. Also, for better insight, one should study in detail the unstable periodic orbits and their asymptotic orbits whose behaviour under perturbation is crucial for the system properties.

The other direction is to make the model of the astrophysical system more realistic. Speaking about the galactic nucleus, one could include the central star cluster (at least in terms of a spherical potential), replace the thin ring by a toroid of finite cross-section, or/and possibly try to account for mechanical interaction of the orbiters with the putative gaseous medium. From the point of view of general relativity, it would however be most interesting to include {\em rotation} (which generates dragging effects, no longer compatible with {\em static} solution). Actually, rotation is almost omnipresent in astrophysics and mainly expected in the case of very compact bodies like black holes, and certainly in the case of orbiting matter like that of accretion discs. Unfortunately, it has proved very difficult to extend the scope from static to {\em stationary} axisymmetric exact setting. It is well known that under such symmetries Einstein equations reduce to the Ernst equation which is completely integrable, but explicit solutions of the corresponding boundary-value problem are rather involved \cite{NeugebauerM-03,KleinR-05,Lenells-11} and not friendly as backgrounds for further extensive numerical studies. Although several related algorithms called {\it generating techniques} have been recognized which can in principle provide {\em any} solution with given (two commuting) symmetries (e.g. \cite{BelinskiV-01}), none of those actually considered in more detail turned out to have satisfactory properties (besides basic ones already familiar before, like the Kerr metric). In particular, simple attempts to ``generate" a {\em physically reasonable} metric for a black hole surrounded by an additional matter or field have not been very successful (e.g. \cite{Tomimatsu-84,KroriB-90,ZellerinS-00,Semerak-02,deCastroL-11} and references therein).

To conclude with the main topic of this seminar, let us also notice that in reality the body whose motion is studied differs from the point-test-particle ideal (and thus the motion from a geodesic one): it is likely endowed with spin or even higher moments; the radiation reaction due to its gravitational emission could be taken into account as well as back reaction due to its non-zero effect on space-time geometry; and, needless to say, if the orbiter was charged, it would also be affected by electromagnetic field, if there was any around.
It would be sensible to employ the methods used in the theory of chaos in order to estimate and compare the significance of various such ``perturbations" which affect the motion in real astrophysical situations.

\subsection*{Acknowledgements}

We thank for support the Czech grants GACR 14-10625S (O.S.) and SVV-267301 (P.S.).
O.S. is grateful to Dirk Puetzfeld for invitation to the Bad Honnef conference and to the Wilhelm und Else Heraeus Stiftung for support there.

\bibliographystyle{unsrt}
\bibliography{Semerak_eom_proceedings_2013}

\end{document}